\documentclass[twocolumn,showpacs,aps,prl,superscriptaddress]{revtex4}

\usepackage{graphicx}
\usepackage{dcolumn}
\usepackage{amsmath}
\usepackage{epsfig}

\input pubboard/babarsym
\newcommand{\beq}{\begin{equation}}
\newcommand{\eeq}{\end{equation}}
\newcommand{\bea}{\begin{eqnarray}}
\newcommand{\eea}{\end{eqnarray}}

\newcommand{\DE}{\ensuremath{\Delta E}}

\newcommand{\pvec}{{\bf p}}

\def\Y#1S{{\Upsilon\rm(#1S)}}
\def\Ks{{K^0_{\scriptscriptstyle S}}}

\def\figurebox#1#2#3{%
    \def\arg{#3}%
    \ifx\arg\empty
    {\hfill\vbox{\hsize#2\hrule\hbox to #2{\vrule\hfill\vbox to #1{\hsize#2\vfill}\vrule}\hrule}\hfill}%
    \else
    {\hfill\epsfbox{#3}\hfill}%
    \fi}

\begin{document}

\preprint{\babar-PUB-\BABARPubYear/\BABARPubNumber} 
\preprint{SLAC-PUB-\SLACPubNumber} 

\begin{flushleft}
\noindent \babar-PUB-08/034\\ 
SLAC-PUB-13362\\
\end{flushleft}

\title{
{\large \bf \boldmath
Observation of $B^0 \to \chi_{c0}K^{*0}$ and Evidence for  $B^+ \to \chi_{c0}K^{*+}$}
}

%
\author{B.~Aubert}
\author{M.~Bona}
\author{Y.~Karyotakis}
\author{J.~P.~Lees}
\author{V.~Poireau}
\author{E.~Prencipe}
\author{X.~Prudent}
\author{V.~Tisserand}
\affiliation{Laboratoire de Physique des Particules, IN2P3/CNRS et Universit\'e de Savoie, F-74941 Annecy-Le-Vieux, France }
\author{J.~Garra~Tico}
\author{E.~Grauges}
\affiliation{Universitat de Barcelona, Facultat de Fisica, Departament ECM, E-08028 Barcelona, Spain }
% ...
\collaboration{The \babar\ Collaboration}
\noaffiliation

\date{\today}

\begin{abstract}
We present the observation of the decay  $B^0\rightarrow \chi_{c0}K^{*0}$ as well as evidence of $B^+ \rightarrow \chi_{c0}K^{*+}$, with an 8.9 and a 3.6 standard deviation significance, respectively, using a data sample of 454 million $\FourS \rightarrow$ \BB\ decays collected with the \babar\ detector at the PEP-II $B$ meson factory located at the Stanford Linear Accelerator Center (SLAC). The measured branching fractions are: ${\cal B}$($B^0$ $\rightarrow$ $\chi_{c0}K^{*0})$ = (1.7 $\pm$ 0.3 $\pm$ 0.2) $\times$ 10$^{-4}$ and ${\cal B}$($B^+ \rightarrow \chi_{c0}K^{*+}$) = (1.4 $\pm$ 0.5 $\pm$ 0.2) $\times$ 10$^{-4}$, where the first quoted errors are statistical and the second are systematic. We obtain a branching fraction upper limit of ${\cal B}$($B^+ \rightarrow \chi_{c0}K^{*+})$ $<$ 2.1 $\times$ 10$^{-4}$ at the 90\% confidence level. 

\end{abstract}


\maketitle

Theoretical predictions of branching fractions and rate asymmetries in non-leptonic heavy-flavor meson decays are difficult due to our limited understanding of the process of quark hadronization. In the simplest approximation, weak decays such as $B \to \jpsi K$ arise from the quark-level process \mbox{$b \to \ccbar s$} through a current-current interaction that can be written as \mbox{$[\cbar \gamma^\mu(1-\gamma_5)c][\sbar \gamma_\mu(1-\gamma_5)b]$}, where $\gamma_{\mu}$ are Dirac matrices ( $\mu$ = 0,1,2,3), $\gamma_5$ = $\gamma_0\gamma_1\gamma_2\gamma_3$ and $\cbar, c, \sbar, b$ are quark spinor fields. The colorless current \mbox{$\cbar \gamma^\mu (1-\gamma_5)c$}, which can create the \jpsi, can  also create the $P$-wave state \chicone. It cannot, however, create the \chiczero, \chictwo or $h_c$ states, so their appearance would have to be explained by a more complex hypothesis. A theoretical prediction can be obtained with the factorization hypothesis~\cite{Bauer:1986bm}, assuming that the weak decay matrix element can be described as a product of two independent hadronic currents. Under the factorization hypothesis, $\B\to\ccbar K^{(*)}$ decays are allowed when the \ccbar pair hadronizes to \jpsi, \psitwos or \chicone, but suppressed when the \ccbar pair hadronizes to \chiczero~\cite{Suzuki:2002sq}. In lowest-order Heavy Quark Effective Theory, the decay rate to the scalar \chiczero is zero due to charge conjugation invariance~\cite{hcGudrun}.

The decay $\Bp\to\chiczero\Kp$ has been observed by Belle and \babar\ with an average branching fraction (${\cal B}$) of $(1.40^{+0.23}_{-0.19}) \times 10^{-4}$~\cite{pdg}, using  \chiczero decays to $\Kp\Km$ or $\pip\pim$. This result is of the same order of magnitude as the branching fraction of the decay $\Bp\to\chicone\Kp$, $(4.9\pm 0.5)\times 10^{-4}$~\cite{pdg}, and is surprisingly large given the expectation from factorization. Using the hadronic \chiczero decays, Belle has obtained an upper limit on $\Bz\to\chiczero\Kz$ of $1.1\times 10^{-4}$ at 90\% confidence level~\cite{belle2007}. No predictions are available for $B$ decays to $\chi_{c0}\Kstar$, so the branching fraction measurement of $\B\to\chi_{c0} K^{*}$ should improve our understanding of the limitations of factorization and of models that do not rely on factorization.

In this paper we report the first observation of $B^0 \rightarrow \chi_{c0}K^{*0}$ and find evidence of the decay $B^+ \rightarrow \chi_{c0}K^{*+}$ ~\cite{cc}. We identify \chiczero mesons through their decays to $h^+h^-$ ($h=K,\pi$), as $\chi_{c0} \to K^+K^-$ and $\chi_{c0} \to \pi^+\pi^-$ have a higher branching fraction than the radiative decay to $\jpsi\g$ ($\jpsi \to l^+l^-, l = \mu$ or $e$), that was used in the previous search for  $B\to\chiczero\Kstar$~\cite{denis}. We identify $K^{*+}$ mesons  through their decay to $\Ks\pi^+$, where $\Ks \to \pi^+\pi^-$, and $K^{*0}$ mesons through their decay to $K^+\pi^-$. 

The data on which this analysis is based were collected with the \babar\ detector~\cite{babar} at the \pep2\ asymmetric-energy \epem\ storage ring. The \babar\ detector consists of a double-sided five-layer silicon tracker, a 40-layer drift chamber, a Cherenkov detector, an electromagnetic calorimeter, and a magnet with instrumented flux return (IFR) consisting of layers of iron interspersed with resistive plate chambers and limited streamer tubes. The data sample has an integrated luminosity of 413~fb$^{-1}$ collected at the $\FourS$ resonance, which corresponds to $(454 \pm 5 )\times 10^6$ \BB\ pairs. It is assumed that the $\FourS$ decays equally to neutral and charged $B$ meson pairs. In addition, 41 fb$^{-1}$ of data collected 40~MeV below the $\FourS$ resonance (off-resonance data) are used for background studies.

Candidate $B$ mesons are reconstructed from five tracks for charged $B$ decays and four tracks for neutral $B$ decays, where three and four tracks, respectively, are consistent with originating from a common decay point within the PEP-II luminous region. Each of the tracks is required to have a transverse momentum greater than 50~\mevc and an absolute momentum less than 10\gevc. The tracks are identified as either pion or kaon candidates, with protons vetoed, using Cherenkov-angle information and ionization energy-loss rate (\dedx) measurements. The efficiency for kaon selection is approximately 80\%, including geometric acceptance, while the probability of misidentification of pions as kaons is below 5\% up to a laboratory momentum of 4\gevc. Muons are rejected using information predominantly from the IFR. Furthermore, the tracks are required to fail an electron selection based on their ratio of energy deposited in the calorimeter to momentum measured in the drift chamber, shower shape in the calorimeter, \dedx, and Cherenkov-angle information. Candidate $\Ks$ mesons are reconstructed from $\pi^+\pi^-$ candidates, and are required to have a reconstructed mass within 15~MeV/$c^2$ of the nominal $K^0$ mass~\cite{pdg}, a decay vertex separated from the $B^+$ decay vertex with a significance of at least five standard deviations, a flight distance in the transverse direction of at least 0.3~cm and a cosine of the angle between the line joining the $B$ and $\Ks$ decay vertices and the $\Ks$ momentum greater than 0.999. 

Four kinematic variables and one event-shape variable are used to characterize signal events. The first kinematic variable, $\DeltaE$, is the difference between the center-of-mass (c.m.)\ energy of the $B$ candidate and $\sqrt{s}/2$, where $\sqrt{s}$ is the total c.m.\ energy. The second is the beam-energy-substituted mass $\mes = \sqrt{(s/2 + \pvec_i \cdot \pvec_B)^2/E_i^2 - \pvec^2_B}$, where  $\pvec_B$ is the reconstructed momentum of the $B$ candidate, and the four-momentum of its parent $\FourS$  in the laboratory frame, ($E_i, \pvec_i$), is determined from nominal colliding beam parameters.  The third kinematic variable is the $K \pi$ invariant mass, $m_{\Kstar}$, used to identify \Kstar candidates, where $K \pi$ is $\Ks \pi^+$ or  $K^+ \pi^-$ for \Kstarp or \Kstarz candidates, respectively. The fourth kinematic variable is the $h^+h^-$ invariant mass, $m_{hh}$, used to identify \chiczero candidates. Candidate  $B$ mesons are required  to satisfy $|\Delta E| <0.1 \gev$, $5.25 <\mes<5.29 \gevcc$, 0.772(0.776) $<$ $m_{\Kstar}$ $<$ 0.992(0.996) GeV/$c^2$ for $B^+$($B^0$) candidates and 3.35 $<$ $m_{hh}$ $<$ 3.50 $\gevcc$ . The event-shape variable is a Fisher discriminant $\mathcal{F}$~\cite{Fisher}, constructed as a linear combination of the absolute value of the cosine of the angle between the $B$ candidate momentum and the beam axis, the absolute value of the cosine of the angle between the thrust axis of the decay products of the $B$ candidate and the beam axis, and the zeroth and second angular moments of energy flow about the thrust axis of the reconstructed $B$.

Continuum quark production ($e^+e^-$ $\rightarrow$ $q\bar{q}$, where $q$ = {\em u,d,s,c}) is the dominant source of background. It is suppressed using another event-shape variable, $|\cos\theta_T|$, which is the absolute value of the cosine of the angle $\theta_T$ between the thrust axis~\cite{thrustaxis} of the selected $B$ candidate and the thrust axis of the rest of the event. For continuum background, the distribution of $|\cos\theta_T|$ is strongly peaked towards 1 whereas the distribution is essentially flat for signal events. Therefore, the relative amount of continuum background is reduced by requiring $|\cos\theta_T| < 0.9$.

Backgrounds from other $B$ meson decays are studied with Monte Carlo (MC) events, using at least $10^3$ times the number of events expected in data for specific decay modes that are the possible sources of background for this analysis.

Potential charm contributions from $B$ $\rightarrow$ $D(\rightarrow K^{*}h^-)h^+$ events are removed by vetoing events with a  reconstructed $K^{*}h^-$ invariant mass in the range 1.83 $<$ $m_{\Kstar h}$ $<$ 1.91~GeV/$c^2$. To remove background from \Dz mesons, a veto is applied to any $K\pi$ pair with an invariant mass in the range 1.83 $<$ $m_{K\pi}$ $<$ 1.91~GeV/$c^2$ for each $B \rightarrow \chiczero (\to h^+h^-)K^{*}$ decay. Studies of MC events show that the largest remaining charmed backgrounds are $B^+$ $\rightarrow \Dzb(\rightarrow \Ks\pi^+\pi^-)\pi^+$ and $B^0$ $\rightarrow D^-(\rightarrow K^+\pi^-\pi^-)\pi^+$, with 12\%  and 10\% passing the veto, respectively. Surviving charmed events have a reconstructed $D$ mass outside the veto range as a result of using a $\pi$ or $K$ candidate that is incorrectly selected from the other $B$ decay in the event. 

A fraction of signal events has more than one $\B$ candidate reconstructed. For those events, the candidate with the highest $\chi^2$ probability of the fitted $B$ decay vertex is selected. Studies of MC events show that less than 11\% of events are reconstructed from the wrong candidate, where these incorrectly reconstructed events are modeled in the fit to data.

After applying all selection criteria, there are five main categories of background from $B$ decays: two- and three-body decays proceeding via a $D$ meson; non-resonant $B \to K^{*}h^+h^-$ and $B \to K\pi\chiczero$;  combinatorial background from three unrelated particles ($K^{*}h^+h^-$); two- or four-body $B$ decays with an extra or missing particle and three-body decays with one or more particles misidentified. Along with selection efficiencies obtained from MC simulation, existing branching fractions for these modes \cite{hfag,pdg} are used to estimate their background contributions that are included separately and fixed in fits to data. For the non-resonant backgrounds, where there is no branching fraction information, fits to sideband data (0.996 $<$ $m_{\Kstar}$ $<$ 1.53 \gevcc and 3.2 $<$ $m_{hh}$ $<$ 3.35 \gevcc) are performed to estimate the background contributions.

In order to extract the signal event yield for the channel under study, an unbinned extended maximum likelihood fit is used. The likelihood function for $N$ events is

\begin{equation}
  \label{eq:Likelihood}
  \mbox{$\mathcal{L}$} \,=\, \frac{1}{N!}\exp\left(-\sum_{i=1}^{M} n_i\right)\, \prod_{j=1}^N 
\,\left(\sum_{i=1}^M n_{i} \, P_{i}(\vec{\alpha},\vec{x_j})\right) ~,
\end{equation}

\noindent where $M=3$ is the number of hypotheses (signal, continuum background, and $B$ background), $n_i$ is the number of events for each hypothesis determined by maximizing the likelihood function, and $P_{i}(\vec{\alpha},\vec{x_j})$ is a probability density function (PDF) with the parameters $\vec{\alpha}$ and variables $\vec{x}$ = (\mes, \DE, $\mathcal{F}$, $m_{\Kstar}$ and $m_{hh}$). The PDF is a product $P_{i}(\vec{\alpha},\vec{x}) = P_{i}(\vec{\alpha}_{\mes} ,\mes ) \times P_{i}(\vec{\alpha}_{\DE} ,\DE ) \times  P_{i}(\vec{\alpha}_{\mathcal{F}}, \mathcal{F}) \times  P_{i}(\vec{\alpha}_{m_{\Kstar}}, {m_{\Kstar}}) \times  P_{i}(\vec{\alpha}_{m_{hh}}, {m_{hh}})$. Studies of MC simulation show that correlations between these variables are small for the signal and continuum background hypotheses. However, for $B$ background, correlations of a few percent are observed between \mes and \DE, which are taken into account by forming a 2-dimensional PDF for these variables. 

The parameters for signal and $B$ background PDFs are determined from MC simulation. All continuum background parameters are allowed to vary in the fit, in order to help reduce systematic effects from this dominant event type. Sideband data, defined to be in the region $0.1 < \Delta E <0.3 \gev$ and $5.25<\mes<5.29 \gevcc$, as well as off-resonance data, are used to model the continuum background PDFs. For the \mes\ PDFs, a Gaussian distribution is used for signal and a threshold function~\cite{argus} for continuum background. For the \DE\ PDFs, a sum of two Gaussian distributions with distinct means and widths is used for the signal and a first-order polynomial for the continuum background. A two-dimensional (\mes, \DE\ ) histogram is used for $B$ background. The signal, continuum and $B$ background $\cal{F}$ PDFs are described using a sum of two Gaussian distributions with distinct means and widths. For $m_{\Kstar}$ PDFs, a sum of a relativistic Breit--Wigner function~\cite{pdg} and a first-order polynomial describes each of the signal, continuum, and $B$ background distributions. Within the $m_{\Kstar}$ fit range, there is also the possibility of $B$ background contributions from non-resonant and higher $K^*$ resonances; these contributions are modeled in the fit using the LASS parameterization~\cite{aston,latham}. The contribution from this background is estimated by extrapolating a $K\pi$ invariant mass projection fitted in a higher-mass region (0.996 $<$ $m_{\Kstar}$ $<$ 1.53 \gevcc) into the signal region. This estimated background is modeled in the final fit to the signal region and assumes there are no interference effects between the $K\pi$ background and the $K^{*}(892)$ signal. Finally, for $m_{hh}$ PDFs, a sum of a relativistic Breit--Wigner function and a first-order polynomial is used to describe signal and a first-order polynomial to describe the continuum and $B$ background distributions. The non-resonant $h^+h^-$ background is modeled by a first-order polynomial, and the background is estimated by extrapolating the invariant mass projection fitted in the lower mass region (3.2 $<$ $m_{hh}$ $<$ 3.35 \gevcc) into the signal region. The signal first-order polynomial component of the $m_{\Kstar}$ and $m_{hh}$ PDFs is used to model misreconstructed events; for example where a $K^+$ from the $K^{*0}$ is reconstructed as a $\chiczero$ daughter particle, and vice versa.

To extract the $B \to \chiczero \Kstar$ branching fractions, ${\cal B}$,  the following equation is used:

\begin{equation}
\label{eq:bf}
\mathcal{B} \,=\ \frac{n_{\rm{sig}}}{N_{BB} \times \epsilon \times {\cal B}({\chi_{c0} \to h^+h^-})}~,
\end{equation}

\noindent where $n_{sig}$ is the number of signal events fitted, $\epsilon$ is the signal efficiency obtained from MC and $N_{BB}$ is the total number of \BB\ events. The efficiencies take into account both ${\cal B}(K^{*+} \to K^0\pi^+) = 2/3$ and ${\cal B}(K^{*0} \to K^+\pi^-) = 2/3$, assuming isospin symmetry, as well as ${\cal B}(K^0 \to \Ks) = 1/2$ and ${\cal B}(\Ks \to \pi^+\pi^-)$\cite{pdg}. The branching fractions are calculated taking into account ${\cal B}(\chi_{c0} \to K^+K^-)$ = $(5.5 \pm 0.6) \times 10^{-3}$ and ${\cal B}(\chi_{c0} \to \pi^+\pi^-)$ = $(7.3 \pm 0.6) \times 10^{-3}$~\cite{pdg}.

\begin{figure}[htb] 
\resizebox{\columnwidth}{!}{
\includegraphics{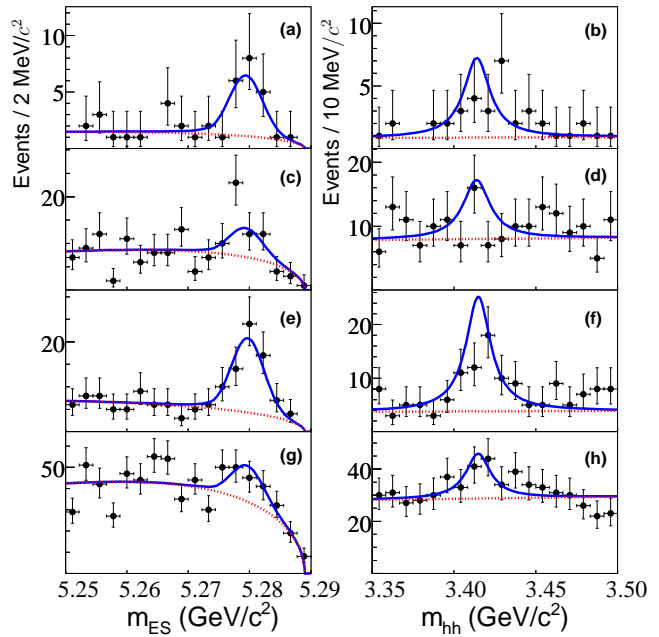}
}
\caption{Maximum likelihood fit projections of  \mes (left column) and $m_{hh}$ (right column) for signal-enhanced samples of $B \to \chiczero \Kstar$ candidates. The dashed line is the fitted background PDF while the solid line is the sum of the signal and background PDFs. The points indicate the data. The plot shows projections for $B^+$ $\to$ $\chi_{c0}K^{*+}(\chi_{c0} \to K^+K^-$) (a) and (b), for $B^+$ $\to$ $\chi_{c0}K^{*+}(\chi_{c0} \to \pi^+\pi^-$) (c) and (d), for $B^0$ $\to$ $\chi_{c0}K^{*0}(\chi_{c0} \to K^+K^-$) (e) and (f), and $B^0$ $\to$ $\chi_{c0}K^{*0}(\chi_{c0} \to \pi^+\pi^-$) (g) and (f).
}\label{fig:fitproj}
\end{figure}

\begin{figure}[htb]
\resizebox{\columnwidth}{!}{
\includegraphics{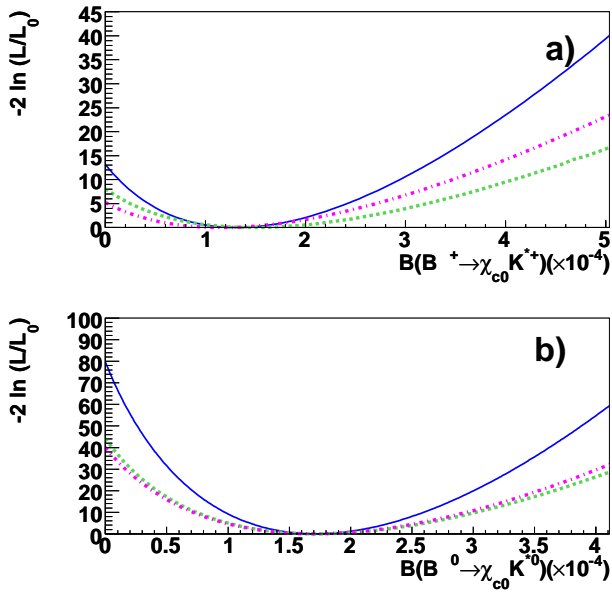}
}
\caption{ Distribution of $-2\ln{\cal L}$ as a function of branching fraction for $B^+$ $\to$ $\chi_{c0}K^{*+}$ (a) and $B^0$ $\to$ $\chi_{c0}K^{*0}$ (b). In each case, the upper dashed line is the decay $\chi_{c0} \to K^+K^-$ and the lower dashed line is the decay $\chi_{c0} \to \pi^+\pi^-$. The solid line is the combination of the two. In all cases systematics contributions are included and the $-2\ln{\cal L}$ distributions have been shifted vertically so the minimum value is 0.
}\label{fig:NLLvBF}
\end{figure}

We observe the decay $B^0$ $\rightarrow$ $\chi_{c0}K^{*0}$ with an 8.9 standard deviation significance and measure the branching fraction ${\cal B}$($B^0$ $\rightarrow$ $\chi_{c0}K^{*0})$ = (1.7 $\pm$ 0.3 $\pm$ 0.2) $\times$ 10$^{-4}$. We find evidence for $B^+ \rightarrow \chi_{c0}K^{*+}$  with a 3.6 standard deviation significance and set a 90\% confidence level upper limit on the branching fraction of 2.1 $\times$ 10$^{-4}$. Figure~\ref{fig:fitproj} shows the fitted \mes and $m_{hh}$ projections for the  $B^+$ $\to$ $\chi_{c0}K^{*+} (\chi_{c0} \to K^+K^-$), $B^+$ $\to$ $\chi_{c0}K^{*+} (\chi_{c0} \to \pi^+\pi^-$), $B^0$ $\to$ $\chi_{c0}K^{*0} (\chi_{c0} \to K^+K^-$)  and $B^0$ $\to$ $\chi_{c0}K^{*0} (\chi_{c0} \to \pi^+\pi^-$) candidates, while the fitted signal yields, measured branching fractions and upper limits are shown in Table \ref{tab:results}. The candidates in Fig.~\ref{fig:fitproj} are signal-enhanced, with a requirement on the probability ratio ${\cal P}_{\mathrm{sig}}/({\cal P}_{\mathrm{sig}} +{\cal P}_{\mathrm{bkg}})$, optimized to enhance the visibility of potential signal, where ${\cal P}_{\mathrm{sig}}$ and ${\cal P}_{\mathrm{bkg}}$ are the signal and the total background probabilities, respectively (computed without using the variable plotted). Figure~\ref{fig:NLLvBF} shows the $-2\ln{\cal L}$ distributions for both $B^+$ $\to$ $\chi_{c0}K^{*+}$ and $B^0$ $\to$ $\chi_{c0}K^{*0}$ as a function of branching fraction. The $-2\ln{\cal L}$ distributions for the final states ($\chi_{c0} \to K^+K^-$ and $\chi_{c0} \to \pi^+\pi^-$) are combined to give final branching fractions shown in Table~\ref{tab:results}. The 90\% confidence level branching fraction upper limit (${\cal B}_{\rm UL}$) is determined by integrating the likelihood distribution (with systematic uncertainties included) as a function of the branching fraction from 0 to ${\cal B}_{\rm UL}$, so that $\int^{{\cal B}_{\rm UL}}_0 {\cal L}\mathrm{d}{\cal B} = 0.9 \int^\infty_0 {\cal L}\mathrm{d}{\cal B}$. The signal significance $S$, in units of standard deviation, is defined as $\sqrt{2\Delta\ln{\cal L}}$, where $\Delta\ln{\cal L}$ represents the change in log--likelihood (with systematic uncertainties included) between the maximum value and the value when the signal yield is set to zero.

\begin{table*}[!ht]
\caption{Total number of events in the fit, $B$ background yields ($B$ bkg), signal yields, efficiencies, and branching fractions ${\cal B}$, measured using $B \to \chiczero \Kstar$ events. Fit bias corrections are applied to the signal yields and branching fractions. The first error is statistical and  the second error is systematic. The significance $S$ is shown for $B^+$ $\to$ $\chi_{c0}K^{*+}$ and $B^0$ $\to$ $\chi_{c0}K^{*0}$ and the branching fraction upper limit, ${\cal B}_{\rm UL}$, at the 90\% confidence level is shown for $B^+$ $\to$ $\chi_{c0}K^{*+}$.}\label{tab:results}
\begin{center}
\begin{tabular}{ccccccccc}
\hline
\hline
Mode&
Total& 
$B$ bkg&
Signal&
Signal&
${\cal B}$ &
${\cal B}_{\rm UL}$&
$S$ ($\sigma$)&\\
&
Events&
&
Yield&
Efficiency(\%)&
($\times$ 10$^{-4}$)&
($\times$ 10$^{-4}$)&
\\
\hline
 $B^+$ $\to$ $\chi_{c0}K^{*+}$& & & & & & &  \\
\hline 
$\chi_{c0} \to K^+K^-$&
156& 
8& 
13 $\pm$ 5&
3.2&
1.6 $\pm$ 0.7 $\pm$ 0.2&
&
 \\
$\chi_{c0} \to \pi^+\pi^-$&
1065&
65&
15 $\pm$ 9&
3.8&
1.2 $\pm$ 0.7 $\pm$ 0.2 &
&
&\\
Combined&
&
&
&
&
1.4 $\pm$ 0.5 $\pm$ 0.2&
2.1&
3.6\\
\hline
 $B^0$ $\to$ $\chi_{c0}K^{*0}$& & & & & &  & \\ 
\hline
 $\chi_{c0} \to K^+K^-$&
690&
20&
47 $\pm$ 10&
11.1&
1.7 $\pm$ 0.4 $\pm$ 0.2 &
 &
\\
$\chi_{c0} \to \pi^+\pi^-$ &
4507&
154&
72 $\pm$ 15&
12.8&
1.7 $\pm$ 0.4 $\pm$ 0.2 &
&
\\
Combined&
&
&
&
&
1.7 $\pm$ 0.3 $\pm$ 0.2 &
&
8.9\\
\hline
\hline
\end{tabular}
\end{center}
\end{table*}

\begin{table}[!tb]
\caption{Summary of systematic uncertainty contributions to the branching fraction measurements $B$ $\to$ $\chi_{c0}K^{*}$. Multiplicative and additive errors are shown as a percentage of the branching fraction. The final row shows the total systematic error on the branching fractions.}\label{tab:sys}
\begin{center}
\begin{tabular}{lcccc}
\hline
\hline
\multicolumn{1}{c}{Error} &
\footnotesize{$\chi_{c0}K^{*+}$}&
\footnotesize{$\chi_{c0}K^{*+}$}&
\footnotesize{$\chi_{c0}K^{*0}$}&
\footnotesize{$\chi_{c0}K^{*0}$}\\
\multicolumn{1}{c}{Source}&
\footnotesize{$\chiczero(KK)$}&
\footnotesize{$\chiczero(\pi\pi)$}&
\footnotesize{$\chiczero(KK)$}&
\footnotesize{$\chiczero(\pi\pi)$}\\
\hline
\multicolumn{2}{l}{Multiplicative errors (\%)}&
&
 \\
Interference&
7.2&
8.3&
6.8&
10.1\\
Tracking&
4.0&
4.0&
3.2&
3.2\\

$\Ks$ Efficiency&
1.7&
1.7&
-&
-\\
Particle ID&
1.9&
2.7&
2.4&
3.2\\
${\cal B}(\chi_{c0} \to h^+h^-)$&
10.9&
8.2&
10.9&
8.2\\
No. of \BB\ &
1.1&
1.1&
1.1&
1.1\\
\hline
Tot. mult.(\%) &
13.9&
12.8&
13.8&
13.8\\
\hline
\multicolumn{2}{l}{Additive errors (\%)}
&
&
 \\
Fit Bias&
1.3&
4.4&
1.8&
3.9\\
$B$ background&
0.5&
4.5&
1.4&
1.5\\

PDF params.&
0.6&
3.4&
0.3&
2.6\\
\hline
Tot. add. (\%)&
1.5&
7.2&
2.3&
4.9\\
\hline
\hline
Total (10$^{-4}$)&
0.2&
0.2&
0.2&
0.2\\
\hline
\hline
\end{tabular}
\end{center}
\end{table}

Contributions to the branching fraction systematic uncertainty are shown in Table~\ref{tab:sys}. The presence of a non-resonant $K^+K^-$  and $\pi^+\pi^-$ can give rise to interference effects, resulting in a departure from the $m_{hh}$ PDF used in the fit to data. In order to estimate how much this can affect the extracted yields, the fit is repeated with the inclusion of a PDF describing the interference between the Breit--Wigner and non-resonant amplitudes in the $m_{hh}$ distribution. This shape consists of the squared modulus of the sum of a Breit--Wigner and a constant amplitude, carrying an arbitrary phase difference. The relative weight of these two components and their phase difference are allowed to vary to obtain the best fit. The signal yields derived from this fit are larger than the nominal fit in Table~\ref{tab:results} and the difference from the nominal fit is used as an estimate of the systematic error in Table~\ref{tab:sys} due to neglecting interference effects. Interference effects between the $K^{*}(892)$ and spin-0 final states (non-resonant and $K^{*}_0(1430)$) integrate to zero if the acceptance of the detector and analysis is uniform; the same is true of the interference between the $K^{*}(892)$ and spin-2 final states ($K^{*}_2(1430)$). Studies of MC events show the efficiency variations are small enough to consider these interference effects insignificant.  The integrated interference between $K^{*}(892)$ and other spin-1 amplitudes such as $K^{*}(1410)$ is in principle non-zero, but in practice is negligible due to the small branching fraction of $K^{*}(1410$) $\to K^+\pi^-$ (6.6 $\pm$ 1.3\%~\cite{pdg}) and the fact that the $K\pi$ mass lineshapes have little overlap. Errors due to tracking efficiency, $\Ks$ reconstruction efficiency and particle identification are assigned by comparing control channels in MC simulation and data. The branching fraction error of $\chiczero \to h^+h^-$ is taken from the combination of previous measurements~\cite{pdg}. The number of \BB\ events is determined with an uncertainty of 1.1\%. To estimate errors due to the fit procedure, 500 MC samples containing the numbers of signal and continuum events measured in data and the estimated number of exclusive $B$ background events are used. The differences between the generated and fitted values are used to estimate small fit biases (see Table~\ref{tab:sys}) that are a consequence of correlations between fit variables. These biases are applied as corrections to obtain the final signal yields, and half of the correction is added as a systematic uncertainty. The uncertainty of the $B$ background contribution to the fit is estimated by varying the known branching fractions within their errors. Each background is varied individually and the effect on the fitted signal yield is added in quadrature as a contribution to the uncertainty.  The uncertainty due to PDF modeling is estimated by varying the PDFs by the parameter errors. In order to take correlations between parameters into account, the full correlation matrix is used when varying the parameters. All PDF parameters that are originally fixed in the fit are then varied in turn, and each difference from the nominal fit is combined in quadrature and taken as a systematic contribution.  

In summary, we have observed the decay $B^0$ $\to$ $\chi_{c0}K^{*0}$ with an 8.9 standard deviation significance and find evidence for $B^+$ $\to$ $\chi_{c0}K^{*+}$ with a 3.6 standard deviation significance, placing an upper limit on the branching fraction. The $B^0$ $\to$ $\chi_{c0}K^{*0}$ branching fraction does not agree with the zero value expected from the color-singlet current-current contribution alone, and is approximately half the $B^0$ $\to$ $\chi_{c1}K^{*0}$ branching fraction ((3.2 $\pm$ 0.6) $\times$ 10$^{-4}$~\cite{pdg}), which is surprising when taking into account factorization expectations.

We are grateful for the excellent luminosity and machine conditions
provided by our \pep2\ colleagues, 
and for the substantial dedicated effort from
the computing organizations that support \babar.
The collaborating institutions wish to thank 
SLAC for its support and kind hospitality. 
This work is supported by
DOE
and NSF (USA),
NSERC (Canada),
CEA and
CNRS-IN2P3
(France),
BMBF and DFG
(Germany),
INFN (Italy),
FOM (The Netherlands),
NFR (Norway),
MIST (Russia),
MEC (Spain), and
STFC (United Kingdom). 
Individuals have received support from the
Marie Curie EIF (European Union) and
the A.~P.~Sloan Foundation.


\begin{thebibliography}{99}
\bibitem{cc}
The use of charge-conjugate modes is implied throughout this paper unless otherwise noted.
\end{thebibliography}
\end{document}